\documentclass[review]{elsarticle}

\usepackage{lineno,hyperref}
\usepackage[font=small,labelfont=bf]{caption}
\usepackage{subcaption}
\usepackage{float}

\journal{Journal of \LaTeX\ Templates}

\usepackage{geometry}
\begin{document}

\begin{frontmatter}

\title{The replacement system of the JUNO liquid scintillator pilot experiment at Daya Bay}
\tnotetext[mytitlenote]{Fully documented templates are available in the elsarticle package on \href{http://www.ctan.org/tex-archive/macros/latex/contrib/elsarticle}{CTAN}.}

\author[mymainaddress,mysecondaryaddress,mythirdaddress]{Wenqi Yan}


\author[mymainaddress,mysecondaryaddress]{Tao Hu}
\author[mymainaddress,mysecondaryaddress]{Li Zhou}
\author[mymainaddress,mythirdaddress]{Jun Cao}
\author[mymainaddress,mysecondaryaddress]{Xiao Cai}
\author[mymainaddress,mysecondaryaddress]{Jian Fang}
\author[mymainaddress,mysecondaryaddress]{Lijun Sun}
\author[mymainaddress,mysecondaryaddress]{Boxiang Yu}
\author[mymainaddress,mysecondaryaddress]{Xilei Sun}
\author[mymainaddress,mysecondaryaddress]{Zeyuan Yu}
\author[mymainaddress,mysecondaryaddress]{Yayun Ding}
\author[mymainaddress,mysecondaryaddress]{Mengchao Liu}
\author[mymainaddress,mysecondaryaddress]{Xiaoyan Ma}
\author[mymainaddress,mythirdaddress]{Xiaohui Qian}
\author[mymainaddress,mysecondaryaddress]{Wanjin Liu}
\author[mymainaddress,mysecondaryaddress]{Yuguang Xie\corref{mycorrespondingauthor}}

\cortext[mycorrespondingauthor]{Corresponding author}
\ead{ygxie@ihep.ac.cn} 

\address[mymainaddress]{Institute of High Energy Physics, Chinese Academy of Sciences, Beijing 100049, China}
\address[mysecondaryaddress]{State Key Laboratory of Particle Detection and Electronics, Beijing 100049, China}
\address[mythirdaddress]{University of Chinese Academy of Sciences, Beijing 100049, China}

\begin{abstract}
The Jiangmen Underground Neutrino Observatory (JUNO), a multi-purpose neutrino experiment, will use 20 kt liquid scintillator (LS). To achieve the physics goal of determining the neutrino mass ordering, 3$\%$ energy resolution at 1 MeV is required. This puts strict requirements on the LS light yield and the transparency. Four LS purification steps have been designed and mid-scale plants have been built at Daya Bay. To examine the performance of the purified LS and find the optimized LS composition, the purified LS was injected to the antineutrino detector 1 in the experimental hall 1 (EH1-AD1) of the Daya Bay neutrino experiment. To pump out the original gadolinium loaded LS and fill the new LS, a LS replacement system has been built in EH1 in 2017. By replacing the Gd-LS with purified water, then replacing the water with purified LS, the replacement system successfully achieved the designed goal. Subsequently, the fluorescence and the  wavelength shifter were added to higher concentrations via the replacement system. The data taken at various LS compositions helped JUNO determine the final LS cocktail. Details of the design, the construction, and the operation of the replacement system are reported in this paper.
\end{abstract}

\begin{keyword}
\texttt{replacement system \sep liquid scintillator \sep JUNO \sep Daya Bay}
\end{keyword}

\end{frontmatter}


\section{Introduction}
Over the recent decades, liquid scintillator (LS) has been used as the target in several neutrino experiments, such as KamLAND\cite{kamland},  Borexino\cite{1}, Daya Bay\cite{2}, Double Chooz\cite{3}, and Jiangmen Underground Neutrino Observatory (JUNO)\cite{4}. Motivated by the physics goals, the requirements on LS are also increased, such as longer attenuation length (A.L.), higher light yield, and ultra-low radioactive background. In the last decade, numerous experiments have been conducted to characterize the performance of LS-based detectors\cite{5,6,A,B,C}.

JUNO is a multi-purpose neutrino experiment\cite{12,yellowbook}, whose primary motivation is to determine the neutrino mass hierarchy and precisely measure the oscillation parameters by detecting reactor antineutrinos. The JUNO experiment is located at about 53 km from the Yangjiang and the Taishan nuclear power plants\cite{13} with a vertical of approximate 700 m. The central detector consists of a 35.4-m acrylic sphere filled with 20-kton LS, viewed by 18,000 20-inch photomultiplier tubes (PMTs) installed on a huge stainless steel (SS) support structure. The central detector is immersed in a water pool to reduce the natural radioactive background from surrounding materials. Liquid scintillator consists of linear alkyl benzene (LAB), 2,5-diphenyloxazole, (PPO; fluor), and p-bis-(o-methylstyryl)-benzene (bis-MSB; wavelength shifter)\cite{14}. Since 2013, many R$\&$D efforts on LS have been carried out. Several purification methods were designed and tested in laboratory. Subsequently, a mid-scale LS plant was built in the experiment hall 5 of the Daya Bay experiment. Twenty tons purified LS were produced. To test the LS optical quantities and natural radioactivity levels, the 20 t Gd-LS in Daya Bay AD1 was drained out and the new LS was filled.  Each AD in Daya Bay consists of a cylindrical target volume with 20 tons of Gd-LS, 192 8-inch PMTs, three automatic calibration units, and multiple layers of shielding and veto detectors\cite{AD1}. A total of eight ADs are utilized at Daya Bay. To avoid the contamination from the Gd-LS, a novel method is that the purified water, which is difficult to dissolve with LS or Gd-LS, was used to replace Gd-LS or LS. Such an operation was realized by a LS replacement system built in Daya Bay EH1. In this paper, the replacement system, an important component of this LS experiment, is introduced, and some test results are presented. 


\section{The JUNO LS pilot experiment}
\begin{figure}[htb]
	\centering 
	\includegraphics[scale=0.55]{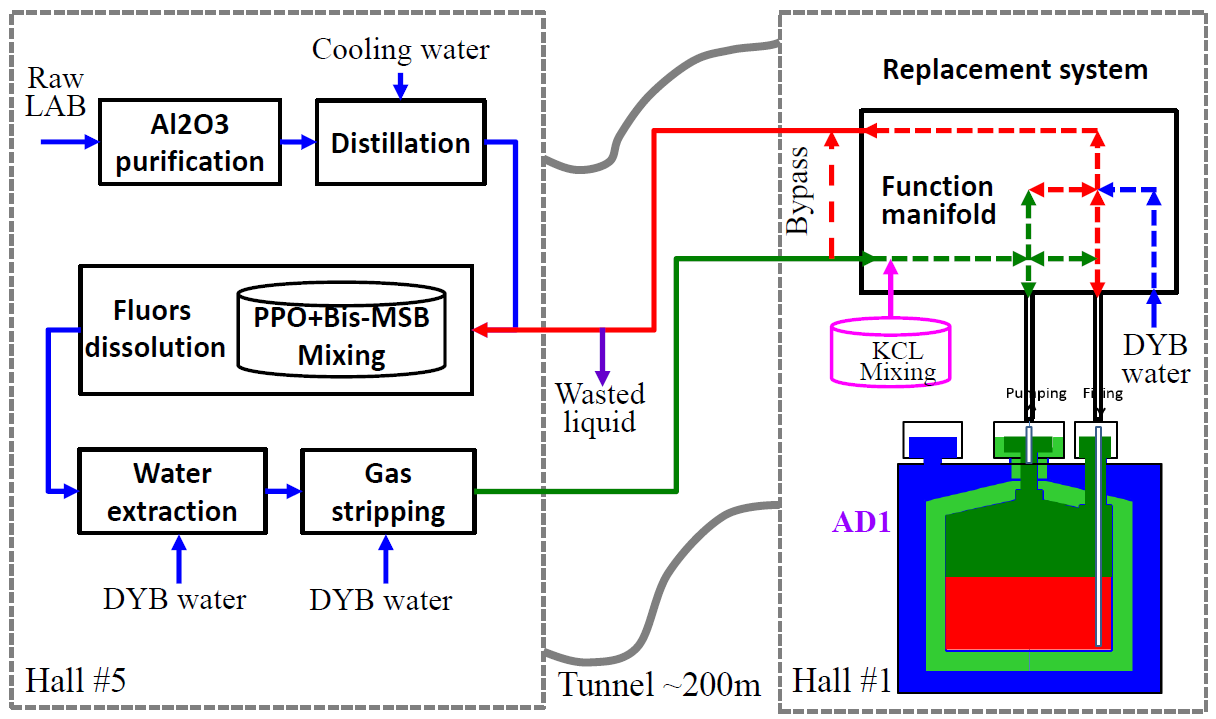}
	\caption{\label{fig:overview} Schematic of the pilot plant experiment}
\end{figure}

The LS pilot experiment mainly aims to examine the purification, which includes Al$_2$O$_3$ column purification, distillation\cite{17}, water extraction, and gas stripping, i.e., the full chain of JUNO LS purification, as shown in Fig~\ref{fig:overview}. In addition, the high purity nitrogen and high purity water (HPW) supply plants are utilized for reducing radon, uranium and thorium contaminations. After those processes of LS pilot, the purified LS is sent to AD1 through the replacement system, which facilitates LS replacement, potassium chloride (KCl) mixing and LS cycling. The connection with AD1 ensures safe operation of the system. 

All the purification plants are placed in experimental hall 5 (EH5) at the Daya Bay site, while the replacement system is built in EH1. The distance between EH1 and EH5 is about 200 m, which leads to a long connecting pipe, thermal dissipation along the line and a delayed response between the purification plants and AD1. It requires the replacement system should have the timely response and temperature adjustment. 

\section{Replacement system}
The replacement system is a ``bridge" connecting purification systems and AD1, which requires complex functions, precise liquid level control and low leakage rate. The specifications of the replacement system are listed in Table~\ref{bs2}. The replacement system is designed to perform several operations that are necessary to fulfill the goals of pilot plant experiment: KCl mixing before draining the Gd-LS, LS replacement, self-circulation, and full-circulation. The KCl doped LS (produced by BNL) was added into AD1 via the self-circulation model. It was used to study the liquid motions during the self-circulation, as well as the energy response calibration of AD1. To avoid the mixture of the Gd-LS in AD1 and the new LS, the Gd-LS was replaced by purified water. Subsequently, the water was replaced by the new LS. Via the full circulation, liquids in the AD can be drained and sent to the facilities in the LS hall for further processing, while newly produced liquids or the re-processed ones were returned to the inner 3-m diameter acrylic vessel (IAV). In the way, the LS properties with different compositions, and the purification efficiencies can be studied. All these functions are realized by the replacement system, which are described in the following sections.
\begin{table}[hbp]
\caption{Parameters of the replacement system}
\centering
\begin{tabular}{ccc}
	\hline
	Parameter & Design & Achieved  \\
	\hline
	Flow rate &	0-500 L/h &	500 L/h, 300 L/h (typical)\\
	Liquid level precision&	$<$ 2 mm&	0.5 mm\\
	Leakage rate&	$<$ 5.0$\times10^{-7}$ mbar L/s&	7.0$\times10^{-8}$ mbar L/s\\
	Surface roughness (Ra)&	$<$ 0.2 $\mu$m &	0.2 $\pm$ 0.1 $\mu$m \\
	Suction head&	4 m @ 500 L/h	&3.9 m @ 473 L/h\\
	\hline       
\end{tabular}
\label{bs2}
\end{table}

\subsection{ LS replacement by water}
The immiscibility and fast separation of LS (mainly LAB) and water are well known. These features not only benefit LS purification such as water extraction but also LS replacement exploiting the density difference. The densities of the Daya Bay Gd-LS and the new LS are both 0.86 g/ml, which is 14$\%$ lower than that of pure water (1 g/ml). This means that the water always stays below the LS and the old Gd-LS with high radioactive background can be completely drained out by filling the AD1 with purified water from the bottom and removing Gd-LS from the top. The complete removal of old Gd-LS is crucial for measuring the radiopurity of the new LS.

\begin{figure}[htb]
	\centering 
	\includegraphics[scale=0.5]{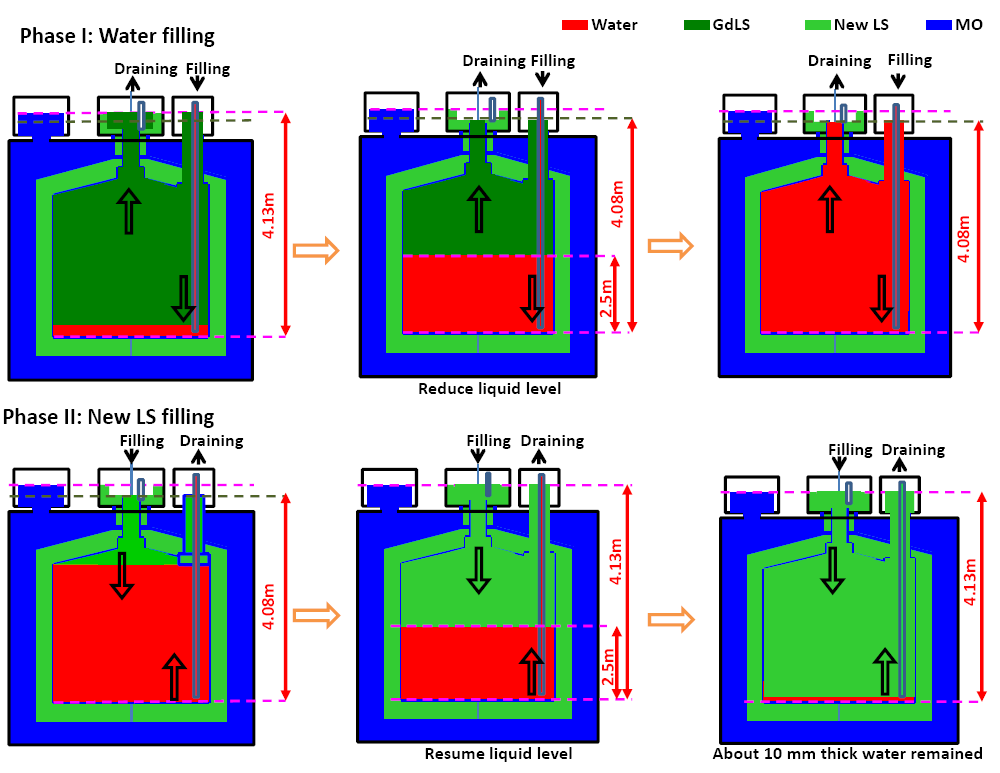}
	\caption{\label{fig:ls replacement process} Procedures involved in LS replacement: filling HPW and draining Gd-LS (top); filling new LS and draining pure water (bottom row).}
\end{figure}

Fig~\ref{fig:ls replacement process} illustrates the procedure of LS replacement in two phases: draining Gd-LS and filling water in Phase I, draining water and fill new LS in Phase II. The time interval between these two phases must be minimized to reduce the risk of damaging the IAV. During phase I, HPW is filled into AD1 by a long pipe at the side, and the flow rate can reach to 500 L/h. To reduce the total stress of IAV, when water is filled to 2.5 m height, the liquid level is reduced to the bottom of the overflow tank. When the interface of Gd-LS and water is close to the neck of IAV according to the estimated accumulated volume, the flow rate must be adjusted to a very low level of about 50 L/h because the neck connection between IAV and overflow tank has a diameter of only 55 mm and length of 820 mm. A high flow rate can lead to a rapid change of the pressure of IAV, which is very risky during the replacement. When the mixture of LS and water is observed at the drain, the flow rate can be restored to its normal value, and filling of water is continued until Gd-LS is completely drained. 

In phase II, the maximum flow rate is only 100 L/h, which is limited by the LS production rate of the pilot plants. The filling and draining ports must be exchanged. The central and side ports are used to fill new LS from the top and drain water from the bottom, respectively. Initially, the flow rate also needs to be controlled at a low level until the oil-water interface is below the neck of IAV. When nearly 40$\%$ water is removed, the liquid level in the overflow tank can be restored to the normal level. During the filling of new LS, its temperature should be controlled. The gas striping plant can adjust the LS temperature to 25$^{\circ}$C, but above 200-m pipe transmission, the temperature can reduce to 15$^{\circ}$C. Therefore, the replacement system is designed with a temperature compensating system based on a heating belt. The water replacement is stopped as soon as the LS/water interface is detected in the draining pipe. The designed draining pipe has an unavoidable gap of 10 mm from the bottom of IAV. Consequently, a small layer of water is left inside the detector.

\subsection{Safety estimation of IAV}
\begin{figure}[htb]
	\centering 
	\includegraphics[scale=0.7]{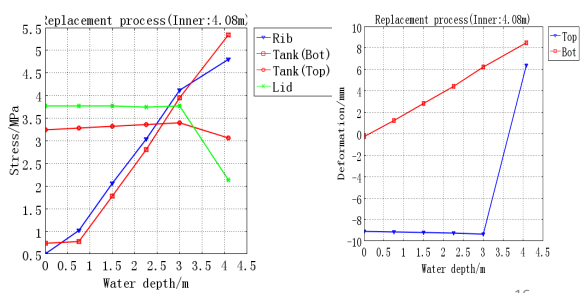}
	\caption{\label{fig:stress_calculation} Variation in the calculated stress and deformation with respect to different water levels.}
\end{figure}

To realize LS replacement, the crucial issue is the safety of IAV during replacement, especially when the vessel is completely filled with water. Therefore, the liquid level must be reduced to the lowest value. Using ANSYS software, the stress and deformation at different liquid levels and water heights were calculated with the real detector model. As shown in the top panel of Fig~\ref{fig:stress_calculation}, the thickness of IAV is only 10 mm. The inner vessel is surrounded by an outer acrylic vessel (OAV), which is filled with Daya Bay LS without Gd doping. The bottom panel of Fig~\ref{fig:stress_calculation} shows the stress and deformation results from the calculation. The stress in the vessel at the bottom plate and at the reinforcing ribs increases with the rising water level, while the stress at the top and lid slightly decreases when the water height is more than 3 m. Combining the dimensions of IAV and overflow, the total water height should be in the range of 4.07 to 4.13 m. Finally, liquid level of 4.08 m is chosen, which is 10 mm higher than the bottom of overflow tank. According to the design standards of the Daya Bay AD, the acrylic vessel can withstand a stress of 5 MPa for a long time and 8 MPa for a short time\cite{Acrylic}, so it is feasible and safe to fill the water to 4.08 m. In this case, based on the calculation results, the maximum stress is 5.343 MPa, and the maximum deformation is 9.4 mm.

\subsection{Interfaces with AD1}
\begin{figure}[htb]
	\centering 
	\includegraphics[scale=0.45]{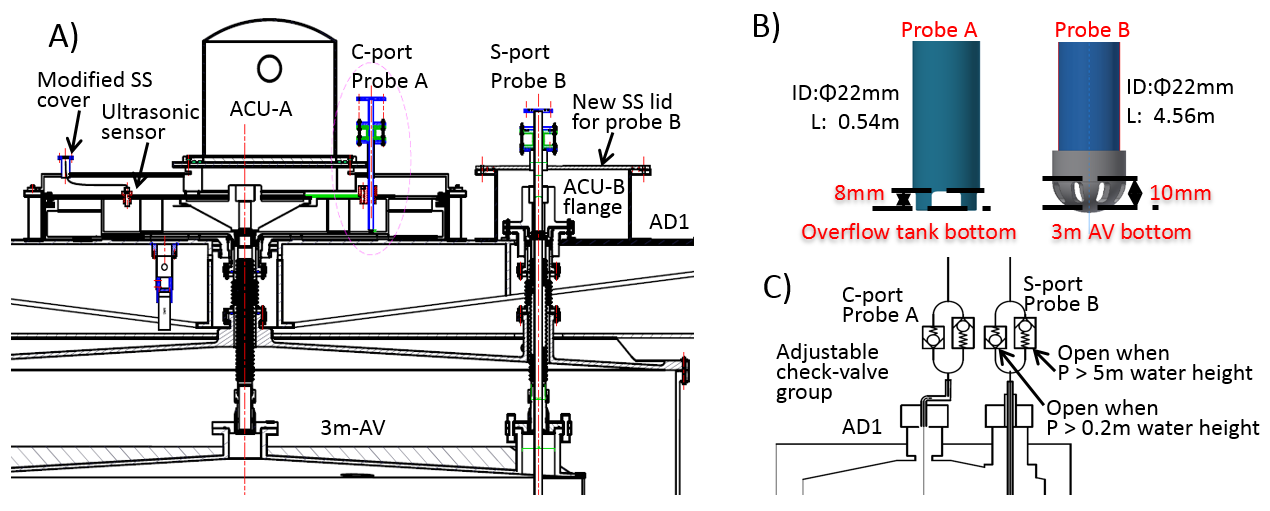}
	\caption{\label{fig:connector} Connectors between AD1 and replacement system.}
\end{figure}

The well-designed AD facilitates the connection of IAV with two ports and monitoring of liquid level in the overflow tank with a high-precision ultrasonic level sensor, as shown in Fig~\ref{fig:connector} (A). By modifying the main SS cover of AD1 and the supporting flange of auto calibration unit B (ACU-B), the new connections on AD1 can meet all the functional requirements of the replacement system. To keep the original monitoring status and not damage the running of AD1, a new ultrasonic sensor with a precision of 0.09 mm was mounted on the lid of the overflow tank of IAV, and signal is sent to the programmable logic controller (PLC) of the replacement system via cables through a newly added flange on the SS cover and an 8-m long bellow, which leaves the water pool and isolates the water. Between the bellow and flange, a customized feedthrough is placed to shield radon. Further, two customized probes have been fabricated. One short probe (Probe-A, 0.538 m) connects the central port (C-port) to the bottom of overflow tank of IAV, and the other long probe (Probe-B, 4.567 m) connects the side port (S-port) to the bottom of IAV. The Probe-A and top part of Probe-B are made of SS, and the long pipe of Probe-B is made of acrylic for light transparency and low radioactivity. Besides the material, the heads of the probes need to be considered carefully, as shown in Fig~\ref{fig:connector} (B). For Probe-A, the head is designed with four 8-mm notches, so this probe can directly contact the bottom of overflow tank, which allows the reduction of liquid level to 10 mm in the overflow tank. For Probe-B, the head must be a cap to avoid direct impact on the 10-mm thick acrylic bottom. The cap has six holes of 6-mm height on the side for allowing liquid flow, and it is kept at a gap of 5 mm from the acrylic bottom to avoid the impact of vibration during filling. Consequently, about 10-mm water can not be drained in the bottom of IAV. 

In addition, since the top of AD1 is 3.6 m lower than the pumps of replacement system, two adjustable check-valve groups are designed for the side and central pipes, as shown in Fig~\ref{fig:connector} (C). Each group consists of a top-open check valve with open pressure higher than 0.2-m water height and a bottom-open check valve with open pressure higher than 5-m water height. Consequently, each pipe prevents the liquid backflow and allows easy extraction of liquid.

\subsection{Operation modes}
\begin{figure}[htb]
 \centering 
 \includegraphics[scale=0.95]{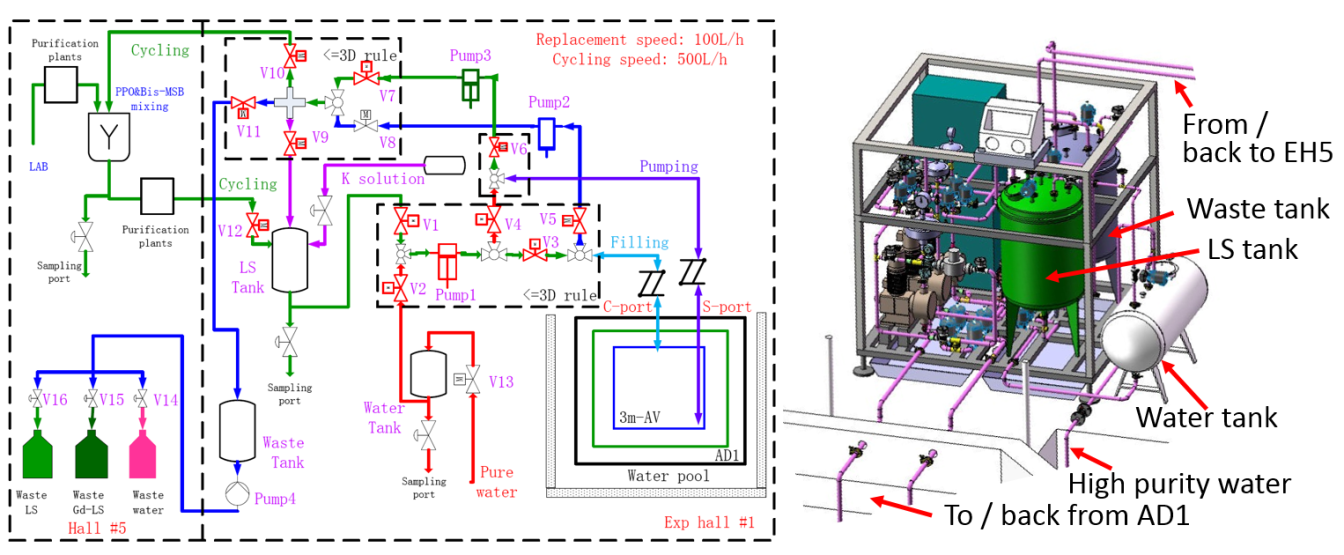}
 \caption{\label{fig:schematic_diagram} Schematic (left) and 3D layout (right) of the replacement system.}
\end{figure}

As mentioned above, four operation modes: self-circulation, water filling, LS filling, and full-circulation, need to be realized. The left plot in Fig~\ref{fig:schematic_diagram} shows the schematic drawing of the replacement system. The different combinations of the four lines can perform all the operations by switching the pumps and valves. To precisely control the operation of the entire system, three diaphragm metering pumps are chosen. Each pump is in continuously variable transmission (CVT). In each line, there are two pumps: one pump matches the flow rate set by the user, and the other pump is automatically adjusted to maintain the liquid level in the overflow tank set by the user. The dynamic balance can be automatically realized by PLC, which is based only on the real-time feedback of the liquid level value. All the plant components (valves, pumps, etc.) are chosen to minimize the leakage rate, electromagnetic interference, and radon contamination.

The system requires three tanks: the LS tank, the water tank, and the waste tank. All the tanks are made of 316L SS for LS compatibility, and the inner surfaces of tanks are electro polished to achieve a surface roughness (Ra) of less than 0.4 $\mu$m. Fig~\ref{fig:schematic_diagram} (right) shows a three-dimension (3D) layout of the entire system and the interfaces with AD1, new LS, HPW and KCl mixing. The pipes and cables are connected to AD1 through two holes drilled on the concrete edge of the water pool. These two holes are sealed after installation to avoid leakage of radon into the water pool.

\begin{figure}[htb]
	\centering 
	\includegraphics[scale=0.5]{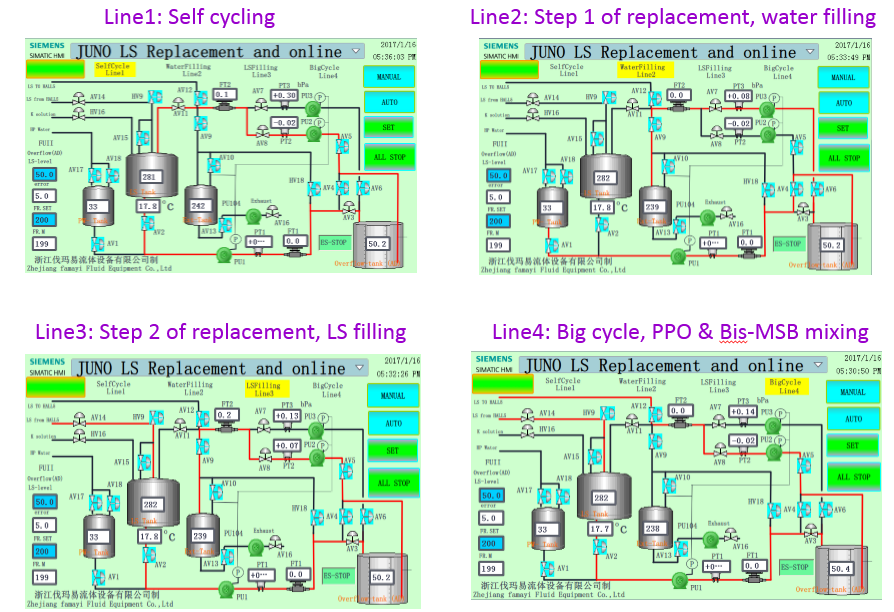}
	\caption{\label{fig:line1_4} Different operation modes at PLC interface.}
\end{figure}

Besides the hardware equipment, the control system is extremely important. Based on Siemens configuration software and PLC, both local (touch screen) and remote (PC) control software are developed to realize all the required functions, including control of operation modes, setting and monitoring of operation parameters, safety interlocks, history curves, alarm, emergency stop, etc. Fig~\ref{fig:line1_4} shows the local control software and the four operation modes (lines), each line can quickly start by one button and run automatically after all the configurations are ready. The details of Line 1 to Line 4 are described as follows (red line indicates current flow path): 

\paragraph{Line 1: Self circulation} In this process, self-circulation of Gd-LS or LS in AD1 and KCl mixing can be realized. As shown in Figure 6 (upper-left), LS in AD1 is drained by pump 2 to the LS buffer tank, then filled back to AD1 via the pump 1. During the circulation, the KCl solution can be mixed into AD1 in a sealed glove box. The KCl mixing test was only performed before the Gd-LS replacement to evaluate the liquid dynamics in the detector.
\paragraph{Line 2: Water filling, step 1 of replacement} In this process, HPW filling and Gd-LS draining are conducted simultaneously. The HPW from EH5 initially enters to the water tank via a polytetrafluoroethylene (PTFE) pipe and is then filled into AD1 from the S-port by PU1. At the same time, Gd-LS is drained from the overflow tank via C-port by PU3 to the waste tank. Finally, Gd-LS is transferred to a liquid bag in EH5 by pneumatic pump (PU4) using a dedicated PTFE pipe, as shown in Figure 6 (upper right).
\paragraph{Line 3: LS filling, step 2 of replacement} This process is basically opposite to Line 2. Initially, the purified LS composed by LAB + 0.5 g/L PPO without bis-MSB from EH5 enters the LS tank and is then filled into AD1 from the C-port by PU1. The waste-water is drained from the bottom of IAV via S-port by PU2 to the waste tank and then sent to the waste pool in EH5 by PU4, as shown in Figure 6 (lower-left). Gd-LS and waste-water share the same PTFE pipe back to EH5 and move to different containers by a manifold. Liquid samples can be taken to check the composition and properties of the mixture offline.

\paragraph{Line 4: Full-circulation and addition of PPO or bis-MSB} In this process, LS is circulated between AD1 in EH1 and the purification plants in EH5 via two $\sim$200-m SS pipes. The LS from EH5 is injected into the LS buffer tank by a diaphragm pump in EH5. It is then filled into AD1 from the C-port by PU1. The LS of AD1 is removed from S-port by PU2 and directly sent to EH5, as shown in Figure 6 (lower right). During full-circulation, more solvents (PPO, bis-MSB) can be added into LAB by a mixing equipment, and different purification plants can be studied separately. 

\section{Cleaning, onsite installation, and leakage check}
\subsection{Cleaning}
\begin{figure}[htb]
	\centering 
	\includegraphics[scale=0.55]{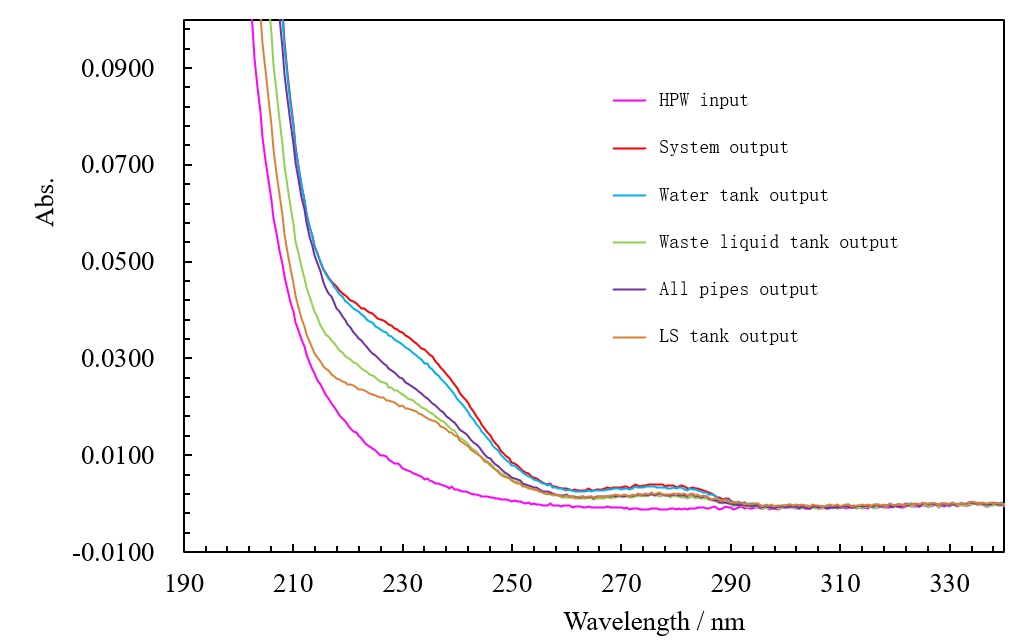}
	\caption{\label{fig:cleaning_check} UV-vis absorption spectrum of the components and entire system for checking the plant cleanness. For the wavelength $>$ 300 nm, no significant difference is observed.}
\end{figure}

To prevent optical or radioactive pollution to the purified LS, the replacement system must be cleaned before onsite installation. The cleaning process was accomplished by two main strategies. Firstly, all the components (pipes, valves, pumps, tanks) were chosen within sanitary class and were processed by adopting sanitary standards for welding technique and electro-polishing (surface-finishing technique). Secondly, all the components in contact with liquid underwent precision cleaning before assembly and installation. Surface degreasing was carefully performed using Alconox detergent and HPW ($\sim$18 M$\Omega\cdot$cm), while the pickling process was conducted with an aqueous of nitric acid ($<$ 20$\%$) to cyclically clean the entire replacement system after assembly. Finally, the cleanness of the plant was examined by two criteria. Firstly, the resistivity of the rinsing water coming out of the cleaning circuit should be higher more than 10 M$\Omega\cdot$cm based on previous experience. Secondly, the ultraviolet-visible (UV-vis) absorption spectrum of output water should not show much degradation, especially for the longer wavelength, as shown in Fig~\ref{fig:cleaning_check}. After cleaning, all the components of the system were flushed with high purity nitrogen for complete drying and then sealed and packed with clean plastic film.

\subsection{Onsite installation}
After cleaning and passing the onsite review by a demo run, the replacement system was installed in EH1 and connected to AD1 in water pool and to purification plants in EH5. During this period, the data acquisition of Daya Bay experiment was stopped for about two weeks, and the water level of water pool in EH1 was lowered below the AD1 cover. 

\begin{figure}[htb]
	\centering 
	\includegraphics[scale=0.95]{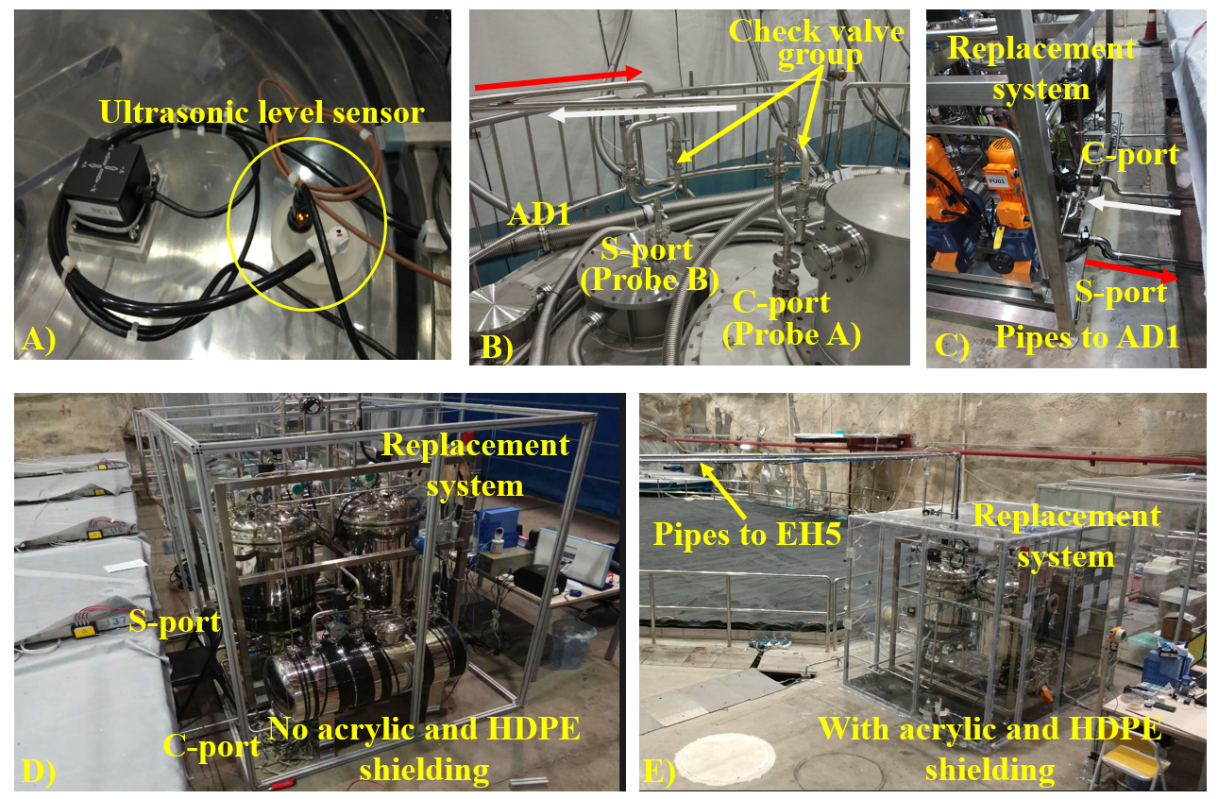}
	\caption{\label{fig:onsite_installation} Onsite installation of replacement system. (a) A new ultrasonic level sensor mounted on the acrylic lid of IAV overflow tank. (b) Probe A and B and the check-valve groups connected to AD1. (c) C-port and S-port of the replacement system. The replacement system with (d) and without (e) sealing by the surrounding acrylic house and high-density polyethylene (HDPE) foil laid on the ground.}
\end{figure}

To mount the probes, the lid of overflow tank (connected with ACU-A) was disassembled and modified with two extra flanges, one for ultrasonic level sensor and the other (C-port) for Probe A. ACU-B was removed and replaced by a new cover with a flange (S-port) to mount Probe B, as shown in Fig~\ref{fig:connector}. Two temporary SS covers were used for shielding AD1 to avoid radon contamination in the 3-day operation. When the two ``new" covers were ready in EH1, one day was spent in mounting the probes and ultrasonic level sensor, sealing the covers and leakage checking, as shown in Fig~\ref{fig:onsite_installation} (a) and (b). 

Further, the replacement system was moved to its assigned position near the water pool in EH1, and the pipes (cleaned and packed in advance) were connected to EH5 and AD1. The pipes connected to EH5 included LS-feed, LS-return, N$_2$ supply, and waste exhaust. The connection between the replacement system and AD1 was realized by only two pipes. These pipes pass-through holes drilled on the cement convex edge and are finally connected to the flanges of Probe A and Probe B with check-valve groups, as shown in Fig~\ref{fig:onsite_installation} (b), (c), and (e). The sensor cables were taken out through a hole sealed with fireproof mud, so the cover of the water pool could be recovered easily. The replacement system after installation is shown in Fig~\ref{fig:onsite_installation} (d). 

\subsection{Radon leakage testing}
One of the goals of the pilot plant experiment is to reduce $^{238}$U contamination of JUNO LS to less than 10$^{-15}$ g/g. Consequently, all the sub-systems must be properly sealed to minimize the radioactive contamination, especially radon. Thus, the diaphragm pumps, valves, and sensors of the replacement system were chosen with a leak rate less than 10$^{-6}$ mbar$\cdot$L/s, and all the tanks were shielded with a dynamic nitrogen blanket. Special attention was paid to the design of flanges with the structures of double O-ring, PTFE gaskets, and KF clamps. The leakage rate of the replacement system was tested with helium, and was found to be lower than 7.0$\times$10$^{-8}$ mbar$\cdot$L/s (Table~\ref{bs2}), which is better than the designed value. 


However, once the LS circulation started, the LS in the IAV was found to be polluted by $^{222}$Rn. The pollution could come from two origins: leakage of the replacement system, or the $^{222}$Rn in the overflow tank of the AD. Thus, some extra measures were taken for the replacement system to reduce the radon contamination. First, an acrylic house was built that surrounded the entire system and was flushed with nitrogen. Second, a layer of HDPE was laid on the ground under the system. Third, some connectors outside the acrylic house were protected by nitrogen boxes.
\begin{figure}[htb]
	\centering 
	\includegraphics[scale=0.6]{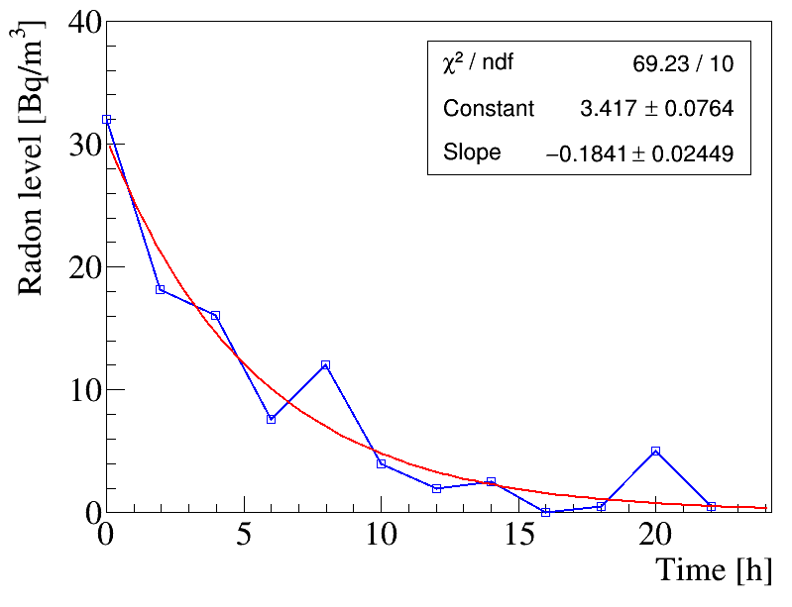}
	\caption{\label{fig:radon_level} Radon level inside the acrylic house filled with the flowing nitrogen gas and laid with HDPE foil on the ground.}
\end{figure}

In the first step, the radon level in the acrylic house increase rapidly to 1000 Bq/m$^3$, because the radon mainly came from the ground under the system. After paving a layer of HDPE, the radon level exponentially decreases to nearly 1 Bq/m$^3$, which is the instrument sensitivity level, as shown in Fig~\ref{fig:radon_level}. The circulation was performed both at the 1000 Bq/m$^3$ and 1 Bq/m$^3$ environment. The radon pollution twas almost a constant in these two stages. It indicated the replacement system was responsible for the radon leakage.

\section{LS replacement progress}
The replacement system was ready on February $\rm4^{th}$, 2017 after onsite review, cleaning, installation and leakage test. From February $\rm4^{th}$ to February $\rm15^{th}$, self-circulation and KCl mixing of the system were completed. In this process, Gd-LS was firstly circulated with a flow rate of 300 L/h, and the combined operation with AD1 and replacement system was examined. Then, 1-L KCl solution was added into the flowing Gd-LS in 1 min to calibrate the $\beta$ spectrum from $^{40}$K. After adding the KCl solution, the GdLS was kept circulated at 300 L/h. The evolution of the spatial distribution of $^{40}$K was used to study the speed of homogenizing the newly added solute. The study found after 7 days of self-circulation, there was still 20$\%$ nonuniformity of $^{40}$K spatial distribution. It means the PPO or bis-MSB can not be added in a short time, otherwise a long circulation time is required to reach an acceptable uniformity. Consequently, in each step of PPO and bis-MSB addition, PPO or bis-MSB was solved in 200 L LS. When the full circulation was started with a 300 L/h circulation speed, the 200 L LS was added slowly in 12 hours. In this way, only 3 days was needed to reach a uniform flour density in the AD. 

\begin{figure}[htb]
 \centering 
 \includegraphics[scale=0.95]{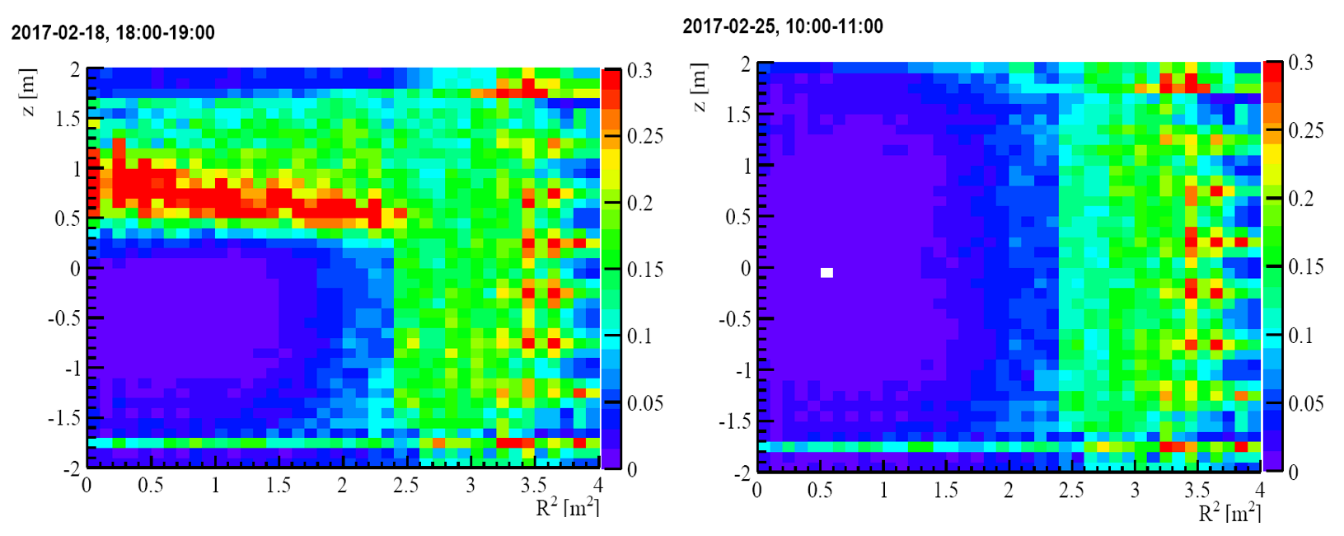}
 \caption{\label{fig:neutrino_events} Event rates in AD1 during (left) and after (right) replacing Gd-LS with water. The region $\rm R^{2}$ $>$ 2.25 m$^2$ is in OAV, which is filled with Daya Bay normal LS, not touched during the replacement.}
\end{figure}

The crucial process, i.e., Gd-LS replacement, started on February $\rm16^{th}$, 2017 and ended on March $\rm7^{th}$, 2017. The first step was water filling. Specifically, HPW from EH5 was filled into AD1, and Gd-LS was removed and sent to liquid bags in EH5 by a 200-m PTFE pipe. As mentioned above, the safety of IAV needed to be ensured, especially when the water level rose to the chimney and the top. A low flow rate of 200 L/h was set, and the rising rate of water level was nearly 28.3 mm/h. The ultrasonic sensor provided the most crucial level value, which was returned to the PLC to adjust the frequency of pumps to maintain stability and balance. At the same time, the detector control system (DCS) of Daya Bay was used to monitor the liquid level for verification. To remotely inspect the running status, several cameras were installed to monitor the pumps, valves, gas bubbles, touch screen, and alarms. It took nearly 5 days to finish the water filling. Subsequently, the second step, i.e., new LS filling, was started. The purified LAB with 0.5 g/L PPO was pumped to the replacement system by a long SS pipe and then filled into AD1. 

Benefiting from the upgraded data acquisition (DAQ) system, the data acquisition for AD1 can be performed independently, so the vertex distribution of natural radioactivity events were monitored during the replacement, as shown in Fig~\ref{fig:neutrino_events}. The range of Z coordinate from -1.5 to 1.5 m and R$^2$ from 0 to 2.25 m$^2$ represents the IAV volume, and the half volume is at Z = 0. Fig~\ref{fig:neutrino_events} (left) shows the vertex distribution after more than half volume of water is filled, therefore events can still be detected in the upper region with Gd-LS, while no events are observed in the lower part with water. After Gd-LS is completely replaced with water, no events can be detected in IAV, as shown in Fig~\ref{fig:neutrino_events} (right). 

After Gd-LS replacement, a series of experiments were conducted, and different LS recipes were examined by adding PPO and bis-MSB in the full-circulation mode from May $\rm 8^{th}$ to July $\rm 27^{th}$, 2017. 
From August, 2017 to the end of 2018, Al$_2$O$_3$ column purification, water extraction, and gas stripping systems were investigated with the replacement system, and the results will soon be reported in a dedicated publication.

\section{Conclusion}
The replacement system has been built, and LS replacement was successfully realized. The replacement system played an important role in the pilot plant experiment of JUNO LS and helped to obtain several crucial results, such as recipe of JUNO LS\cite{bay2020optimization}, water extraction efficiency, stripping efficiency, and radon shielding effect. Further, it can serve as a valuable reference for developing the filling system of 20-kton LS in JUNO experiment.

\section*{Acknowledgements}
This work was supported by the National Natural Science Foundation of China (Grant No. 11390384), Strategic Priority Research Program of the Chinese Academy of Sciences (Grant No. XDA10010500) and the CAS Center for Excellence in Particle Physics (CCEPP). We thank the Daya Bay collaboration for contributing EH1-AD1 for the LS studies, for the excellent design of the antineutrino detector, and for the help during our system design, review and installation. We thank the Daya Bay onsite people for their kind support. We specially thank Paolo Lombardi and all the European members of JUNO-LS group for their useful suggestions and cooperation.

\bibliography{mybibfile}

\end{document}